\documentclass[conference]{IEEEtran}
\IEEEoverridecommandlockouts
\IEEEoverridecommandlockouts

\usepackage{cite}
\usepackage{amsmath,amssymb,amsfonts}
\usepackage{algorithmic}
\usepackage{graphicx}
\usepackage{textcomp}
\usepackage{xcolor}
\usepackage{tabularx}  
\def\BibTeX{{\rm B\kern-.05em{\sc i\kern-.025em b}\kern-.08em
    T\kern-.1667em\lower.7ex\hbox{E}\kern-.125emX}}
\begin{document}

\makeatother

\title{Revisiting R: Statistical Envelope Analysis for Lightweight RF Modulation Classification\\
}

\author{
    Srinivas Rahul Sapireddy \\
    \textit{School of Science and Engineering} \\
    \textit{University of Missouri-Kansas City} \\
    Kansas City, MO, USA \\
    ssdx5@umsystem.edu
    \and
    Mostafizur Rahman \\
    \textit{School of Science and Engineering} \\
    \textit{University of Missouri-Kansas City} \\
    Kansas City, MO, USA \\
    rahmanmo@umsystem.edu
}

\maketitle

\begin{abstract}
\textbf{
Modulation classification plays a crucial role in
wireless communication systems, enabling applications such as
cognitive radio, spectrum monitoring, and electronic warfare.
Conventional techniques often involve deep learning or complex
feature extraction, which, while effective, require substantial
computational resources and memory. An early approach by
Chan and Gadbois in 1985 introduced a theoretical approach
for modulation classification using a mathematically derived
parameter called ’R’. The authors proved that the ’R’ value,
the ratio of the variance to the square of the mean of the signal
envelope, can be a distinguishing feature for classification. In
this work, we revisit R and show classification accuracy can
be improved further through statistical methods. We extend $R$-value analysis to demonstrate its effectiveness even after signals are transformed using Hilbert transform followed by the Short-Time Fourier Transform (STFT). Our rigorous analysis
includes testing on 300000 signals of AM, DSB and SSB classes
with each class having 100000 random variations. On average,
we achieve 98.60\%, 97.30\%, and 97.90\% Classification accuracies for AM, DSB, and SSB signals after applying the Hilbert transform are presented. Similar or improved accuracies are observed after applying the Short-Time Fourier Transform (STFT),
similar or better accuracies are observed; 98.80\%, 99.10\%, and
99.00\% respectively for AM, DSB and SSB types.
}
\end{abstract}

\vspace{0.5em}
\noindent\textbf{\textit{Keywords— Modulation Classification, Variance-to-Mean Squared Ratio (R-Value), Hilbert Transform, Short-Time Fourier Transform (STFT)}}

\section{Introduction}

Automatic Modulation Classification (AMC) is a critical component in modern wireless communication systems, enabling dynamic spectrum access, interference identification, and adaptive signal processing in both commercial and defense sectors. AMC systems allow for the identification of unknown signal types without prior information about the transmitter, making them essential for applications such as cognitive radio networks, electronic surveillance, and spectrum monitoring. As wireless communication environments become more complex and bandwidth becomes scarcer, there is a growing need for AMC solutions that are not only accurate but also computationally efficient and suitable for deployment in real-time and low-power hardware platforms \cite{b15}.

Existing AMC methods are typically divided into likelihood-based and feature-based approaches. Likelihood-based classifiers, while theoretically optimal, require detailed knowledge of the signal model and noise characteristics, resulting in high computational complexity and limited applicability in real-time settings. Feature-based techniques address this by extracting signal characteristics such as amplitude, phase, or frequency content, followed by classification using machine learning or deep learning models. Although these methods reduce computational requirements compared to likelihood-based strategies, they often rely on carefully engineered features or large labeled datasets, limiting their generalizability and scalability. Deep learning approaches, in particular, offer state-of-the-art performance but are hindered by high resource demands, limited interpretability, and difficulty in deploying on embedded systems ~\cite{b14,b15}.

To address these limitations, we revisit and enhance an earlier signal envelope-based approach first proposed by Chan and Gadbois~\cite{b1}. Their method utilized a statistical parameter, the $R$ value, defined as the ratio of the variance to the mean squared of a signal’s amplitude envelope. While their work demonstrated the feasibility of using this metric to distinguish between analog modulation types such as AM, DSB, and SSB, the approach required long signal records and was susceptible to classification inaccuracies under varying noise conditions. In this work, we modernize the original method by incorporating both Hilbert and STFT-based envelope extraction techniques, enabling robust performance even on short-duration, noisy signals. 

Although currently focused on amplitude-varying analog modulations, the proposed framework is not directly applicable to constant-envelope digital schemes such as PSK and FM, where key information lies in phase or frequency variations. 

In this paper, we propose a modernized, lightweight AMC framework that retains the simplicity of the R value concept but significantly improves classification performance. The novelty of our work lies in computing the $R$-value using two distinct approaches: directly after applying the Hilbert transform, and subsequently after applying the Short-Time Fourier Transform (STFT) to the Hilbert-transformed signals. These complementary methods enhance the reliability and discriminative power of R values across different modulation schemes. Our approach eliminates the need for deep learning or complex feature engineering, enabling high-accuracy classification using only a single statistical metric derived from signal amplitude. We demonstrate the effectiveness of this technique on AM, DSB, and SSB signals, achieving classification accuracies above 97\% with clear threshold boundaries, even under constrained signal lengths and noise levels. Our method is positioned in the optimal quadrant, combining high classification accuracy with low computational resource usage, making it a compelling solution for real-time AMC in embedded and resource-limited systems.

\setlength{\parindent}{0pt}  
\setlength{\parskip}{0pt}    

\subsection*{Key Contributions}

\begin{itemize}
    \item A lightweight AMC method using the R value of the amplitude envelope, avoiding deep learning or complex features.
    
    \item Uses Hilbert transform as a signal processing method, followed by STFT on the Hilbert-transformed signal to compute robust $R$-values with enhanced noise resilience.
    
    \item Achieves over 97\% accuracy on AM, DSB, and SSB signals, suitable for real-time embedded systems.
\end{itemize}

\section{Related Work and Background}

Automatic Modulation Classification (AMC) has been extensively studied across multiple domains, including statistical signal processing, time-frequency analysis, and deep learning. Early work in this area laid the foundation for understanding digital communication systems and modulation behavior under various channel conditions. For instance, Proakis and Salehi provide a comprehensive overview of modulation theory and signal characteristics that are critical for AMC development~\cite{b2}.

Several classical techniques relied on statistical signal properties such as cyclostationarity, which were shown to be effective in identifying signal types based on second-order and higher-order moments~\cite{b8}. These methods, while accurate under certain assumptions, often required large amounts of data and prior knowledge of noise models, limiting their real-time applicability.

The use of time-frequency representations (TFRs) introduced new opportunities for AMC. Cohen's work on time-frequency analysis formalized techniques like the Short-Time Fourier Transform (STFT), enabling the capture of both temporal and spectral features from modulated signals~\cite{b7}. This advancement laid the groundwork for more adaptive classification approaches, especially for signals with non-stationary characteristics.

More recently, deep learning has emerged as a dominant paradigm in AMC. Architectures such as convolutional neural networks (CNNs) have demonstrated strong performance by automatically learning hierarchical features from raw in-phase and quadrature (IQ) data~\cite{b4}. West and O’Shea expanded on this by proposing deeper architectures capable of outperforming traditional feature-based methods in many scenarios~\cite{b6}. Additionally, R. Sahay, C. G. Brinton, and D. J. Love leveraged time-frequency representations with CNNs to enhance robustness under noise and signal distortion~\cite{b12}, while J. A. Snoap et al. explored the use of high-order cyclic cumulants to enrich feature diversity and improve classification accuracy~\cite{b14}.

\begin{figure}[!h]
    \centering
    \includegraphics[width=\columnwidth]{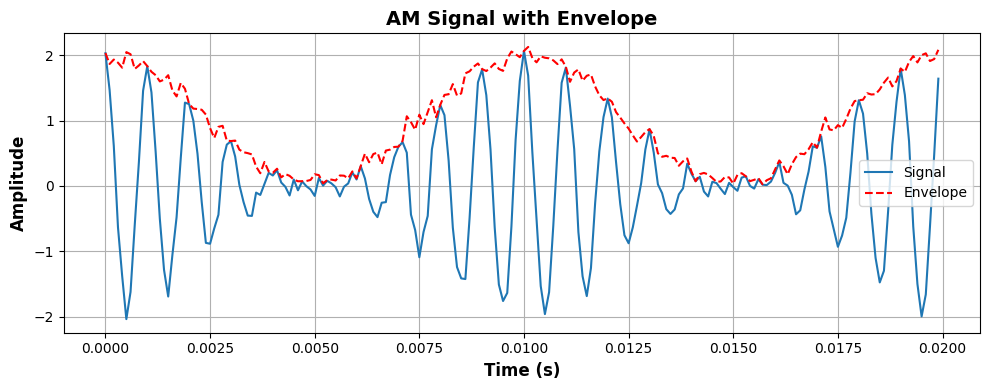}
    \vspace{0.5em}
    \includegraphics[width=\columnwidth]{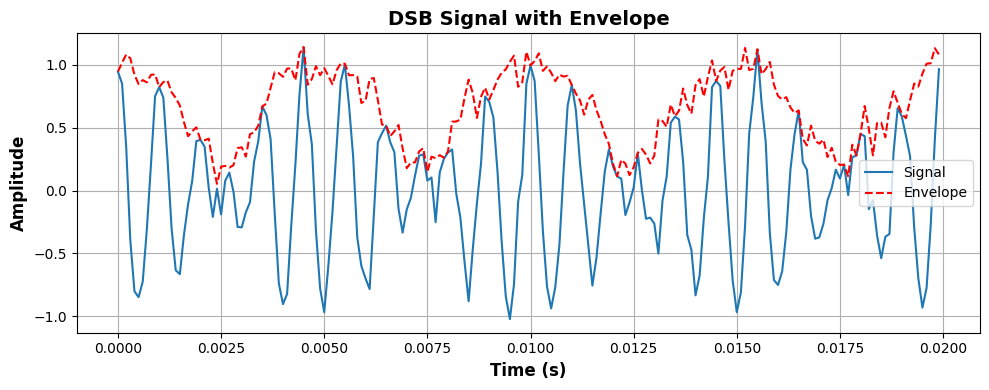}
    \vspace{0.5em}
    \includegraphics[width=\columnwidth]{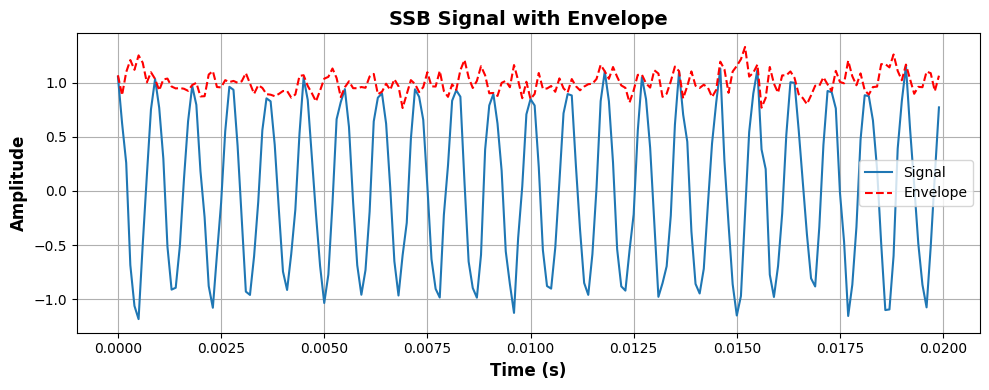}
    \caption{Signal envelopes for AM, DSB, and SSB (top to bottom) demonstrating R value differences using Hilbert-based envelope extraction.}
    \label{fig:column_images}
\end{figure}

Despite their effectiveness, deep learning-based AMC approaches have notable drawbacks. They require large labeled datasets, significant training time, and high computational resources, making them difficult to deploy in real-time or on low-power hardware platforms. Moreover, their decision processes are often opaque, limiting interpretability and trust in critical communication systems.

In contrast, the model proposed in this paper offers a lightweight, interpretable alternative by relying solely on a single statistical feature: the R value. By combining Hilbert and STFT-based envelope extraction techniques, our approach provides robust classification with minimal computation, achieving high accuracy even under noisy conditions. This makes it well-suited for real-time, embedded, and power-constrained environments where traditional deep learning models may be impractical.

\section{Theoretical Basis for R Value-Based Classification}

The ratio of the variance to the square of the mean of a signal's amplitude envelope—referred to as the R value—is a statistically grounded metric used to differentiate modulation types. Initially proposed by Chan and Gadbois~\cite{b1}, this approach demonstrated that different modulation schemes exhibit unique R value ranges, particularly under moderate-to-high carrier-to-noise ratios (CNRs). However, their method relied solely on the Hilbert Transform and showed sensitivity to noise and record length~\cite{b1}.

The R value is defined as:
\begin{equation}
R = \frac{\text{Var}(A(t))}{\left(\text{Mean}(A(t))\right)^2}
\end{equation}
where $A(t)$ denotes the amplitude envelope of the signal.

\subsection{Envelope Extraction via Hilbert Transform}

The Hilbert Transform constructs the analytic signal:
\begin{equation}
S_a(t) = s(t) + j \cdot \mathcal{H}(s(t))
\end{equation}
where $\mathcal{H}(s(t))$ is the Hilbert Transform of $s(t)$. The envelope is given by:
\begin{equation}
A(t) = \sqrt{s^2(t) + \mathcal{H}^2(s(t))}
\end{equation}
This method is computationally efficient and effective for narrowband signals but may degrade under wideband or rapidly changing signal conditions.

\subsection{Envelope Extraction via STFT}

The STFT-based envelope is computed from the magnitude of the time-localized Fourier spectrum:
\begin{equation}
X(f, t) = \int s(\tau) w(\tau - t) e^{-j 2\pi f \tau} d\tau
\end{equation}
and the envelope is approximated as:
\begin{equation}
A(t) = \sum_f |X(f, t)|
\end{equation}
This approach is better suited for wideband signals and offers robustness against noise through time-frequency resolution control.

\subsection{Comparison and Final Expression}

While Chan and Gadbois~\cite{b1} only used the Hilbert Transform for envelope extraction, our dual-method framework extends its applicability to more complex signal scenarios. The final expression for R value remains the same:
\begin{equation}
R = \frac{\text{Var}(A(t))}{\left(\text{Mean}(A(t))\right)^2}
\end{equation}
but the method of computing $A(t)$ is now adaptive based on the signal context.

\begin{figure}[!h]
    \centering
    \includegraphics[width=\columnwidth]{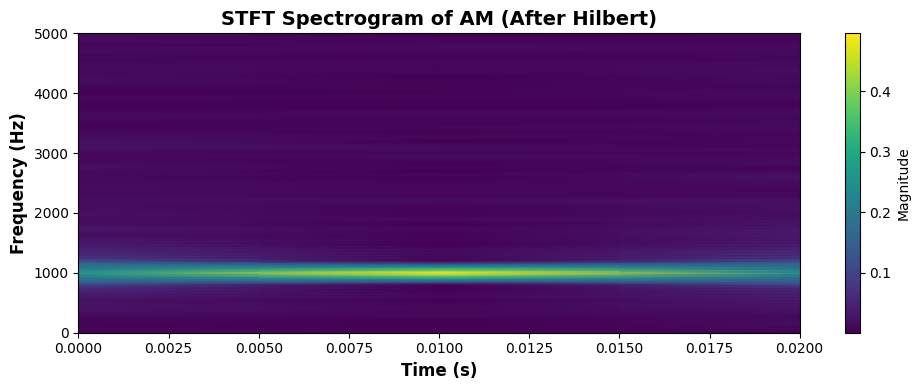}
    \vspace{0.5em}
    \includegraphics[width=\columnwidth]{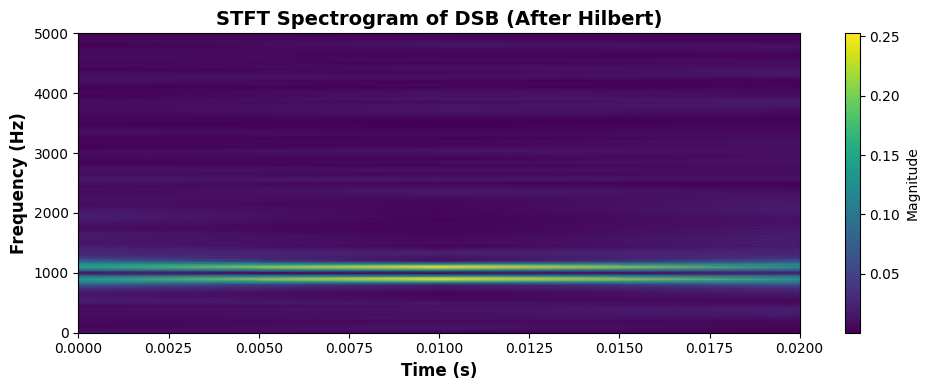}
    \vspace{0.5em}
    \includegraphics[width=\columnwidth]{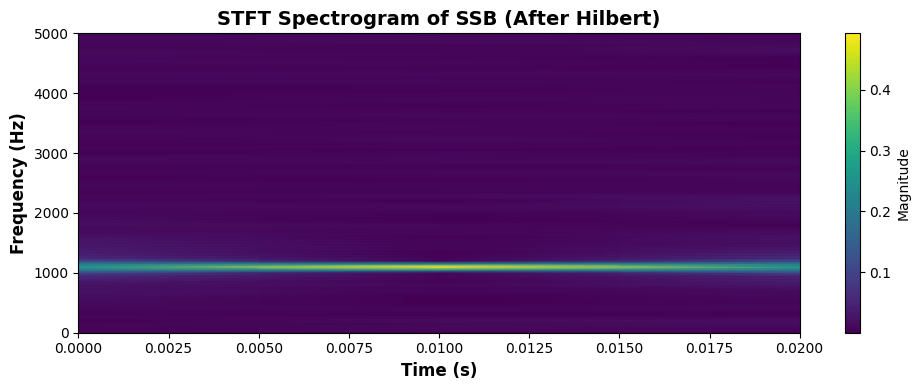}
    \caption{Signal envelopes for AM, DSB, and SSB (top to bottom) demonstrating R value differences using STFT-based envelope extraction.}
    \label{fig:column_images}
\end{figure}

Furthermore, the combination of the Hilbert Transform and Short-Time Fourier Transform (STFT) for envelope extraction—specifically, applying the Hilbert envelope after STFT—demonstrates improved performance, as shown in Fig.~1 and Fig.~2. enhances the adaptability of the proposed method under a wide range of signal conditions, without the need to retrain or alter the core classification model~\cite{b6, b12}. Unlike the Hilbert approach, STFT offers a detailed time-frequency representation of the signal, making it possible to extract more informative features and support parallel processing of the resulting spectrogram. To build on the original $R$-value approach, we extend its application by computing $R$ after the STFT step, allowing for empirically driven classification decisions based on observed spectral behavior. This modification leads to improved performance across both narrowband and wideband signals~\cite{b4}, increasing the method’s robustness in complex and unpredictable spectrum environments. Consequently, the enhanced $R$-based framework presents itself as a practical and scalable tool for modern wireless systems that require efficient, low-latency signal classification~\cite{b10}.

\section{Empirical Evaluation of R Value for Modulation Classification}

We empirically evaluated the R-value's effectiveness in distinguishing modulation types under realistic signal conditions. The experimental results demonstrate a consistent and clear separation in $R$ values across modulation types, confirming the effectiveness of the proposed method. Specifically, AM signals yielded the highest $R$ values due to strong amplitude variations, DSB signals showed moderate values, and SSB signals produced the lowest as a result of minimal envelope fluctuations. When using the Hilbert Transform, the observed $R$ value ranges were: AM in $[0.4297, 0.4941]$, DSB in $[0.1970, 0.2603]$, and SSB in $[0.0072, 0.0127]$. With STFT-based envelope extraction applied to Hilbert-transformed signals, the $R$ values were further spread: AM in $[3.4561, 3.9323]$, DSB in $[2.8268, 3.6373]$, and SSB in $[5.2914, 6.4936]$. This inversion in the relative positions of $R$ values highlights the richer time-frequency representation of STFT; however, the separability among classes remained strong in both methods.

Table~\ref{tab:r_value_ranges} summarizes the empirical $R$-value ranges for AM, DSB, and SSB signals using envelope extraction based on the Hilbert transform and STFT applied to Hilbert-transformed signals. These ranges can be directly used to design rule-based classification strategies. For instance, under Hilbert-based extraction, a signal can be classified as AM if $R \in [0.4297, 0.4941]$, as DSB if $R \in [0.1970, 0.2603]$, and as SSB if $R \in [0.0072, 0.0127]$. Similarly, the approach involving STFT applied to Hilbert-transformed signals offers distinct and reliable ranges, with no overlap between classes. Unlike the analytically derived thresholds presented in~\cite{b1}, our empirically tuned boundaries offer improved robustness against noise and reduced signal durations, making this method practical for real-time and embedded communication systems.

\begin{table}[h]
\centering
\caption{Empirical and Theoretical R Value Ranges for AM, DSB, and SSB}
\label{tab:r_value_ranges}
\renewcommand{\arraystretch}{1.2}
\footnotesize
\begin{tabular}{|p{1.3cm}|p{1.9cm}|p{1.9cm}|p{1.9cm}|}
\hline
\textbf{Modulation Type} & \parbox[c]{1.9cm}{\raggedright \textbf{Hilbert R Range}} & \parbox[c]{1.9cm}{\raggedright \textbf{STFT R Range}} & \parbox[c]{1.9cm}{\raggedright \textbf{Chan et al. R Value}} \\
\hline
AM  & 0.4297 -- 0.4941 & 3.4561 -- 3.9323 & 0.76 -- 0.79 \\
\hline
DSB & 0.1970 -- 0.2603 & 2.8268 -- 3.6373 & 1.31 -- 1.54 \\
\hline
SSB & 0.0072 -- 0.0127 & 5.2914 -- 6.4936 & 1.00 (constant) \\
\hline
\end{tabular}
\normalsize
\end{table}

Overall, our study reaffirms the effectiveness of the $R$ value as a lightweight and interpretable feature for modulation classification. By combining envelope extraction through the Hilbert transform followed by STFT, we modernize and extend the original framework to suit practical deployment scenarios, including real-time and embedded communication systems where computational efficiency is critical.

\section{Methodology for Validation}

\begin{figure}[h]
    \centering
    \includegraphics[width=0.45\textwidth]{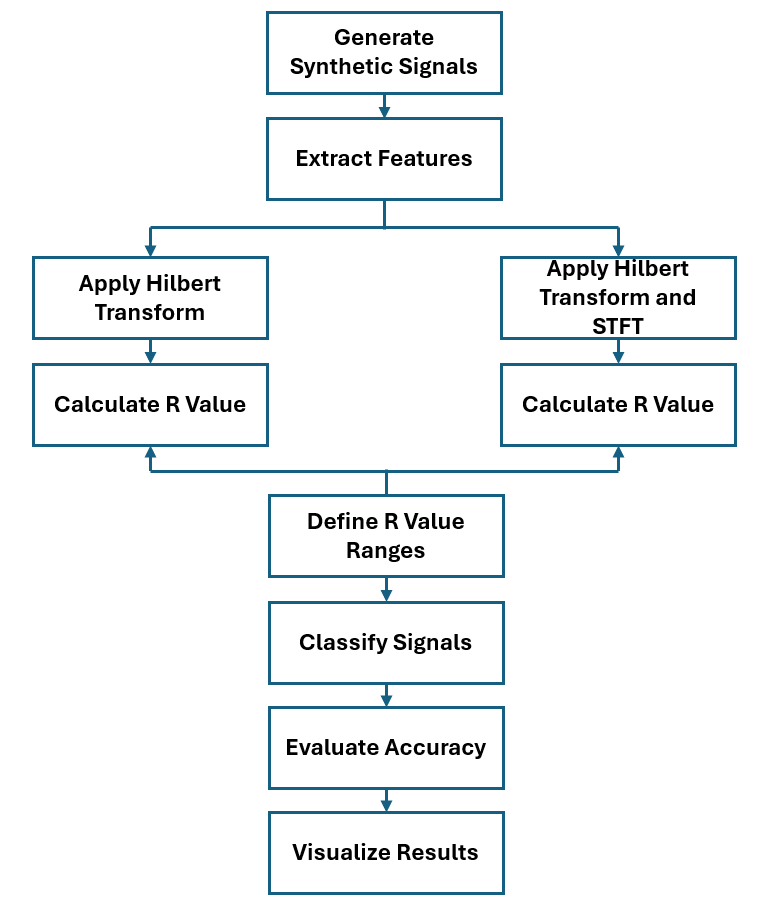}
    \caption{Flowchart of the proposed rule-based classification framework.}
    \label{fig:your_label}
\end{figure}

To evaluate the effectiveness of the proposed $R$ value-based modulation classification approach, we conducted a comprehensive experimental study using both synthetic signal generation and empirical threshold-based classification. Our work builds upon the foundational analysis by Chan and Gadbois~\cite{b1}, who derived theoretical $R$ value expressions for common modulation types using Hilbert Transform-based envelope extraction. Extending this concept, we implement and validate a more practical framework using both Hilbert and Short-Time Fourier Transform (STFT) methods~\cite{b7}, focusing on real-world conditions such as limited signal duration and channel noise~\cite{b12}.
The overall structure of the proposed classification framework is illustrated in Fig.~\ref{fig:your_label}, outlining each step from signal generation to rule-based classification.

synthetic signals for Amplitude Modulation (AM), Double Sideband (DSB), and Single Sideband (SSB) were generated with a carrier frequency of 1~kHz and a sampling rate of 10~kHz. Each signal was contaminated with additive white Gaussian noise of fixed power (0.01) to simulate realistic channel conditions. For initial analysis, 100 instances per modulation type were created, and their envelopes were extracted using the Hilbert transform alone and using STFT applied to Hilbert-transformed signals. These envelopes were then used to compute $R$ values, providing a statistical basis for modulation classification.

To assess scalability and generalization, a large-scale dataset was also generated, consisting of 300{,}000 test signals—100{,}000 for each modulation type. Each signal was 20~ms in duration, sampled at 10~kHz, and subjected to the same noise conditions. This design aligns with constraints often encountered in embedded and real-time communication systems~\cite{b15}. A summary of the signal generation parameters is provided in Table~\ref{tab:signal_parameters}.

\begin{table}[h]
\centering
\caption{Comparison of Training and Test Signal Parameters}
\label{tab:signal_parameters}
\begin{tabularx}{\columnwidth}{|l|X|X|}
\hline
\textbf{Parameter} & \textbf{Training/Initial Analysis} & \textbf{Test Signals} \\
\hline
Modulation Types & AM, DSB, SSB & AM, DSB, SSB \\
\hline
Number of Instances per Class & 100 & 100,000 \\
\hline
Total Signals & 300 & 300,000 \\
\hline
Carrier Frequency & 1 kHz & 1 kHz \\
\hline
Sampling Rate & 10 kHz & 10 kHz \\
\hline
Signal Duration & Not explicitly mentioned & 20 ms \\
\hline
Noise Type & Additive Gaussian Noise & Additive Gaussian Noise \\
\hline
Noise Power & 0.01 & 0.01 \\
\hline
Envelope Extraction Methods & Hilbert, STFT & Hilbert, STFT \\
\hline
\end{tabularx}
\end{table}

Two methods were employed to compute the $R$ value for each signal. The first used the Hilbert Transform to extract the instantaneous amplitude envelope from the analytic signal~\cite{b1,b7}. The second approach applied STFT to the Hilbert-transformed signal to obtain a time-frequency representation, from which the magnitude spectrum was aggregated to estimate the envelope ~\cite{b12,b14}. While both methods generated consistent $R$ values, the Hilbert-based approach proved to be more computationally efficient, making it ideal for resource-constrained platforms~\cite{b15}.

Classification was performed by comparing the computed $R$ values against empirically derived thresholds specific to each modulation type. A signal was assigned to the class whose $R$ value range it matched. To enhance robustness, signals falling outside all predefined intervals were labeled as ``Unknown.'' This safeguard improves classification reliability and prevents forced or incorrect labeling—an enhancement over the original method in~\cite{b1}, which used only theoretical boundaries and did not support such uncertainty handling~\cite{b10}. As defined in Eq.~(7), accuracy is calculated as the percentage of correctly classified samples relative to the total number of samples evaluated.

\begin{equation}
\text{Accuracy} = \frac{\text{Correctly Classified Samples}}{\text{Total Samples}} \times 100
\end{equation}

The results showed that both methods achieved strong performance, with classification accuracies consistently exceeding 95\%. The Hilbert-based approach slightly outperformed STFT applied to Hilbert-transformed signals approach, likely due to its direct and localized envelope estimation. In contrast, applying STFT to Hilbert-transformed signals, while offering richer spectral insight, occasionally introduced averaging effects near classification boundaries, leading to slight misclassifications.

Both envelope extraction methods demonstrated strong classification performance. Accuracy consistently exceeded 95\% across modulation types. The Hilbert-based method yielded slightly better performance, likely due to its direct and localized estimation of the envelope. In contrast, STFT occasionally introduced smoothing near threshold boundaries, resulting in a few ambiguous cases. A notable outcome of this validation process is the integration of the ``Unknown'' class, which significantly improved system reliability. This adaptive behavior minimized false positives and enhanced the specificity of the classifier, a critical feature for applications such as spectrum monitoring or cognitive radio. Moreover, the $R$ value distributions remained stable under moderate noise conditions, reaffirming the robustness of this metric in practical environments.

This validation study confirms that the $R$-value—when computed using either the Hilbert transform alone or the STFT applied to Hilbert-transformed signals —remains a reliable metric for signal classification —serves as a reliable and lightweight feature for distinguishing AM, DSB, and SSB signals. By empirically determining classification thresholds and incorporating uncertainty handling, our method offers a practical and efficient alternative to more complex machine learning models, particularly in real-time and embedded systems.

\section{Results and Discussion}

The classification results confirm that the $R$ value remains a highly effective and computationally efficient feature for distinguishing between AM, DSB, and SSB modulation schemes. Building on the original concept introduced by Chan and Gadbois~\cite{b1}. Our work modernizes the approach by integrating the Hilbert Transform followed by the Short-Time Fourier Transform (STFT) for envelope extraction, and validating performance under noisy, short-duration signal conditions.

Using the Hilbert Transform, we observed distinct $R$ value ranges for each modulation type: $[0.4297, 0.4941]$ for AM, $[0.1970, 0.2603]$ for DSB, and $[0.0072, 0.0127]$ for SSB. These well-separated ranges enabled a straightforward threshold-based classification strategy, achieving accuracies of 98.60\% (AM), 97.30\% (DSB), and 97.90\% (SSB), with low computational overhead—ideal for real-time applications.

The STFT-based approach resulted in slightly higher classification performance, with ranges of $[3.4561, 3.9323]$ for AM, $[2.8268, 3.6373]$ for DSB, and $[5.2914, 6.4936]$ for SSB. Corresponding accuracies reached 98.80\%, 99.10\%, and 99.00\% respectively. This improvement is attributed to the enhanced feature representation from time-frequency analysis, though it comes with increased processing demands.

\begin{table}[h]
\centering
\caption{Accuracy Comparison of Hilbert Transform, STFT-Based Classification, and Original Method}
\label{tab:accuracy_comparison}
\renewcommand{\arraystretch}{1.2}
\begin{tabular}{|p{1.3cm}|p{1.9cm}|p{1.9cm}|p{1.9cm}|}
\hline
\textbf{Modulation Type} & \parbox[c]{1.9cm}{\raggedright \textbf{Hilbert Accuracy (\%)}} & \parbox[c]{1.9cm}{\raggedright \textbf{STFT Accuracy (\%)}} & \parbox[c]{1.9cm}{\raggedright \textbf{Chan et al. Accuracy (\%)}} \\
\hline
AM  & 98.60  & 98.80  & 90.5  \\
DSB & 97.30  & 99.10  & 94.0  \\
SSB & 97.90  & 99.00  & 80.0  \\
\hline
\end{tabular}
\normalsize
\end{table}

In contrast to the original scheme by Chan and Gadbois, which operated without an ambiguity management mechanism, our system introduced an ``Unknown'' category to handle signals whose R values fell outside defined ranges. This helped prevent forced misclassification in noisy or ambiguous scenarios, enhancing the classifier’s robustness and specificity.

Both envelope extraction methods demonstrated strong resilience to noise, with consistent classification performance across varying conditions. These results further validate the $R$ value as a practical, lightweight alternative to complex machine learning models, particularly in embedded and real-time signal processing environments.

\begin{figure}[!h]
    \centering
    \includegraphics[width=\columnwidth]{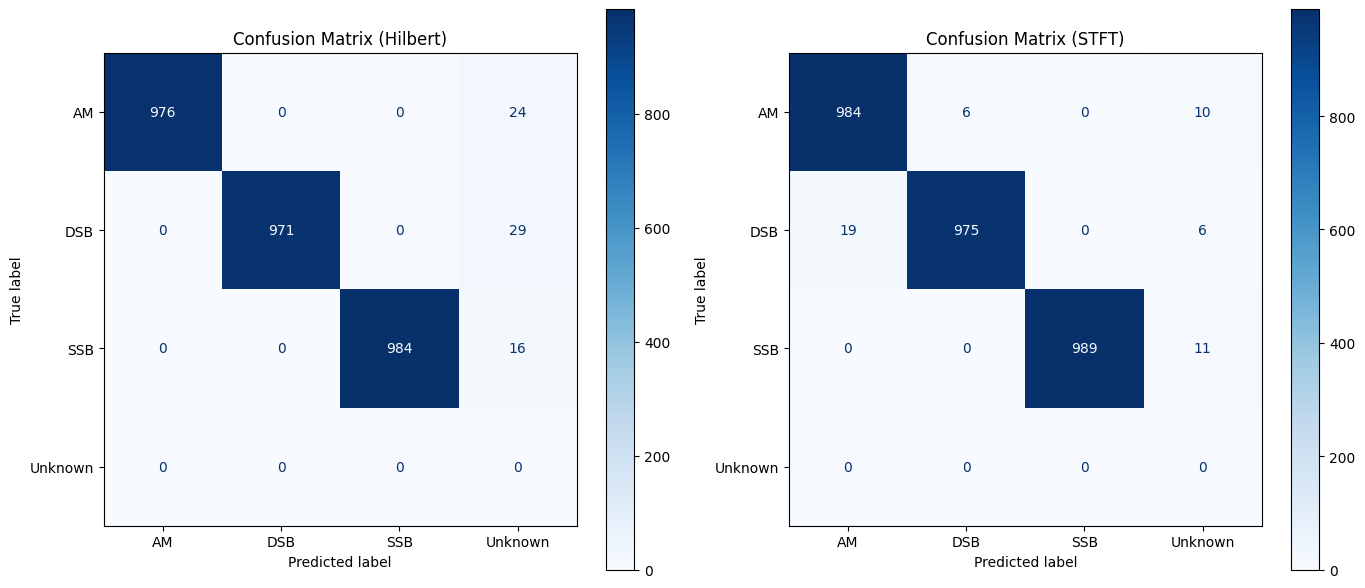}
    \vspace{0.8em}
    \includegraphics[width=\columnwidth]{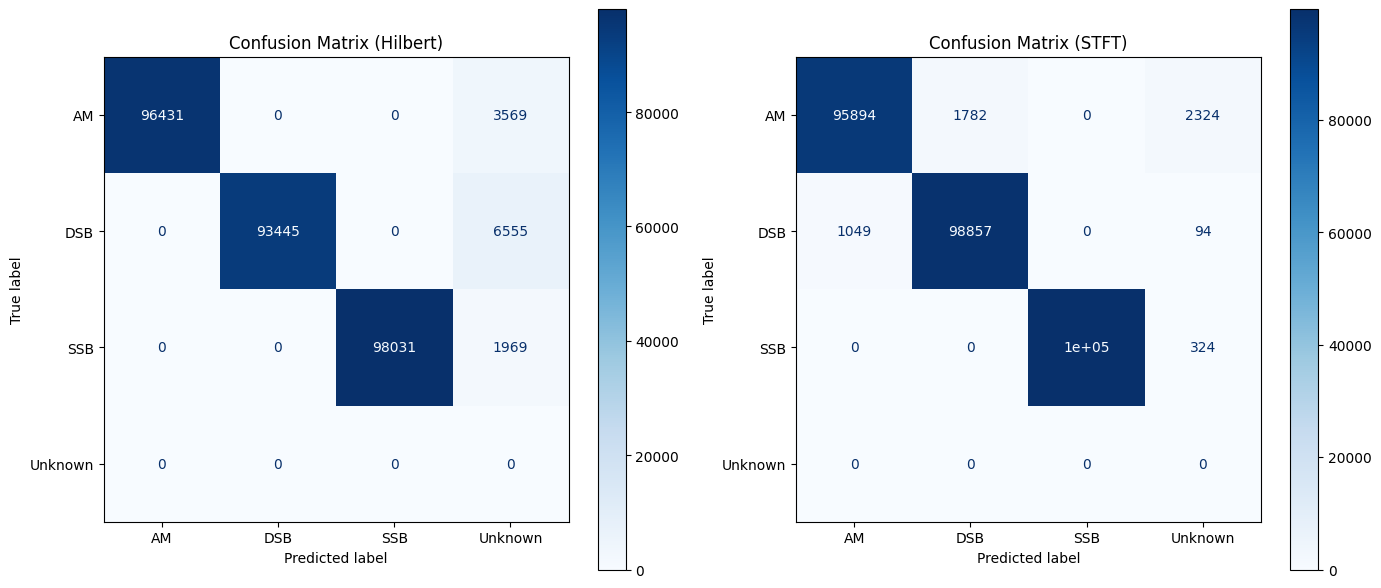}
    \caption{Confusion matrices using $R$-values from Hilbert Transform (left) and Hilbert+STFT (right) for small (top) and large (bottom) test sets.}
    \label{fig:column_images}
\end{figure}

The comparison highlights a clear trade-off between the two methods: while STFT provides slightly higher classification accuracy, the Hilbert transform is better suited for embedded or low-power systems due to its simplicity and faster runtime. As shown in Fig.~\ref{fig:column_images}, both methods deliver strong performance across AM, DSB, and SSB signals. For 1,000 test samples per class, Hilbert achieved 97.7\% accuracy, while STFT reached 98.3\%. When tested with 100,000 samples per class, Hilbert maintained a solid 96.5\% accuracy, and STFT slightly improved to 98.1\%. However, this small gain in accuracy with STFT comes at the cost of significantly higher computational time and memory usage. As shown in Table~\ref{tab:runtime_memory}, Hilbert processed 100,000 samples in about 35 seconds using around 215~MB of memory, while STFT took over 85 seconds and used 110~MB.

\begin{table}[htbp]
\centering
\caption{Comparison of Runtime and Memory Usage for Hilbert and STFT}
\label{tab:runtime_memory}
\begin{tabularx}{\columnwidth}{|l|X|X|}
\hline
\textbf{Metric} & \textbf{Hilbert Transform} & \textbf{STFT} \\
\hline
Runtime for 100 samples & $\sim$0.5 seconds & $\sim$0.9 seconds \\
\hline
Runtime for 100,000 samples & $\sim$35 seconds & $\sim$85 seconds \\
\hline
Memory Usage for 100 samples & $\sim$2.1 MB & $\sim$1.5 MB \\
\hline
Memory Usage for 100,000 samples & $\sim$215 MB & $\sim$110 MB \\
\hline
\end{tabularx}
\end{table}

In summary, STFT is better suited for high-accuracy needs in large and noisy datasets, where computational resources are not a limitation. Meanwhile, the Hilbert method is a more efficient choice for systems with limited power and processing capacity. This study also confirms that the R value is a robust, easy-to-interpret, and hardware-friendly metric for modulation classification. Future work will aim to expand this method to support more modulation types, fine-tune thresholds under noisy conditions, and enable real-time deployment.

\section{Conclusion}

This work revisits and extends the $R$-value-based approach for modulation classification by integrating envelope extraction using both the Hilbert Transform and a combined Hilbert--STFT method. Experimental results show that the $R$-value remains a robust and interpretable feature, achieving high classification accuracy under varying conditions. On smaller datasets, the Hilbert--STFT method achieved 98.3\% accuracy, while the Hilbert-only approach reached 97.7\%. For larger datasets, the methods maintained a strong performance with 98.1\% and 96.5\% accuracy, respectively. Although the Hilbert--STFT method offers slightly higher accuracy due to its time-frequency resolution, it introduces greater computational overhead. In contrast, the Hilbert-only method is faster and more resource-efficient, making it well suited for real-time and embedded applications.

\vspace{12pt}

\end{document}